\documentclass[conference]{IEEEtran} 
\pdfoutput=1

\IEEEoverridecommandlockouts

\usepackage{cite}
\usepackage[english]{babel}
\usepackage{graphicx}
\usepackage[T1]{fontenc}
\usepackage[latin1]{inputenc}
\usepackage[psnames,svgnames]{xcolor}
\usepackage[justification=centering, caption=false]{subfig}
\usepackage{amsmath,amssymb,amsthm,latexsym}
\usepackage{algorithm}
\usepackage{algpseudocode} 
\usepackage{xspace}
\usepackage{cite}
\usepackage[bookmarks=false]{hyperref}

\hypersetup{%
pdftitle={Regenerating Codes: A System Perspective},
pdfauthor={Steve Jiekak, Anne-Marie Kermarrec, Nicolas Le Scouarnec, Gilles Straub, Alexandre Van Kempen},
pdfkeywords={practical, regenerating codes, adaptive, system, implementation},
pdfstartview={FitH},
urlcolor=darkblue,
pdfborder=0 0 0,
bookmarksnumbered
}%

\begin{document}
\sloppy

\renewcommand{\O}[1]{$\mathcal{O}\left(#1\right)$}
\newcommand{\Omc}[1]{$\Omega\left(#1\right)$}
\newcommand{\xor}{{\small \textsf{xor}}\xspace}
\newcommand{\cc}[1]{\mathrm{cc}(#1)}
\def\zin{\mathrm{in}}
\def\zout{\mathrm{out}}
\def\zcoor{\mathrm{coor}}
\def\MC{\mathcal{C}}
\def\MG{\mathcal{G}} 
\def\MM{\mathcal{M}}
\def\MB{\frac{\MM}{k}}
\def\lcm{\mathrm{lcm}}
\def\dopt{d_{\mathrm{opt}}}
\def\nopt{n_{\mathrm{opt}}}

\date{}

\title{Regenerating Codes: A System Perspective\thanks{\hspace{-0.25cm}%
\textbf{Published in ACM SIGOPS Operating Systems Reviews (2013)\cite{OSR2013}.\newline{}
Available at \url{http:/dx.doi.org/10.1145/2506164.2506170}} \newline{}
\textbf{Extended version of a paper accepted at DISCCO 2012\cite{DISCCO2012}.\newline{}Available at \url{http:/dx.doi.org/10.1109/SRDS.2012.58}} \newline{}
\copyright 2012 IEEE. Personal use of this material is permitted. Permission from IEEE must be obtained for all other uses, in any current or future media, including reprinting/republishing this material for advertising or promotional purposes, creating new collective works, for resale or redistribution to servers or lists, or reuse of any copyrighted component of this work in other works.}}

\author{

	\IEEEauthorblockN{Steve Jiekak\IEEEauthorrefmark{1}\IEEEauthorrefmark{2},
                                Anne-Marie Kermarrec\IEEEauthorrefmark{3}, 
				Nicolas Le Scouarnec\IEEEauthorrefmark{1},
				Gilles Straub\IEEEauthorrefmark{1} and
				Alexandre Van Kempen\IEEEauthorrefmark{1}}

	\IEEEauthorblockA{
	\IEEEauthorrefmark{1}Technicolor, Rennes, France}
	\IEEEauthorrefmark{2}EPFL, Lausanne, Suisse\\
	\IEEEauthorrefmark{3}INRIA Rennes - Bretagne Atlantique, France\\

}
\maketitle


\begin{abstract}
The explosion of the amount of data stored in cloud systems calls for more efficient paradigms for redundancy. While replication is widely used to ensure data availability, erasure correcting codes provide a much better trade-off between storage and availability. Regenerating codes are good candidates for they also offer low repair costs in term of network bandwidth. While they have been proven optimal, they are difficult to understand and parameterize. In this paper we provide an analysis of regenerating codes for practitioners to grasp the various trade-offs. More specifically we make two contributions: \emph{(i)} we study the impact of the parameters by conducting an analysis at the level of the system, rather than at the level of a single device; \emph{(ii)} we compare the computational costs of various implementations of codes and highlight the most efficient ones. Our goal is to provide system designers with concrete information to help them choose the best  parameters and design for regenerating codes.
\end{abstract}



\begin{figure*}[t]
\centering
\subfloat[Tradeoff curve]{\centering
        \includegraphics[width=0.28\linewidth]{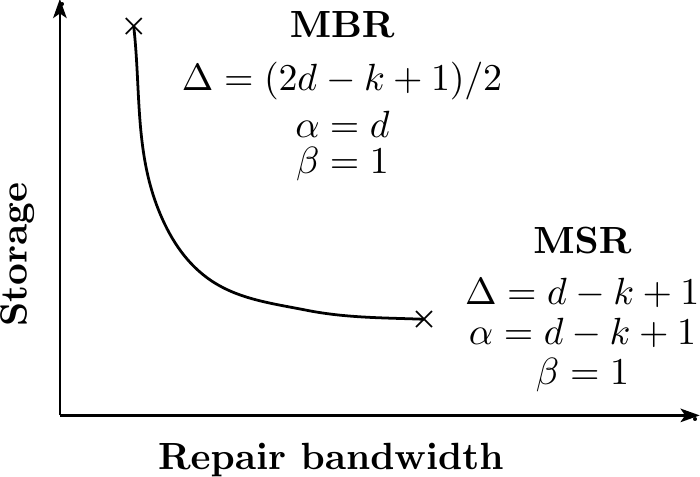}%
        \label{fig:cr}
}\hfill
\subfloat[Optimal repair using regenerating codes)]{\centering
        \includegraphics[height=0.200\linewidth,trim=0 -10 0 0]{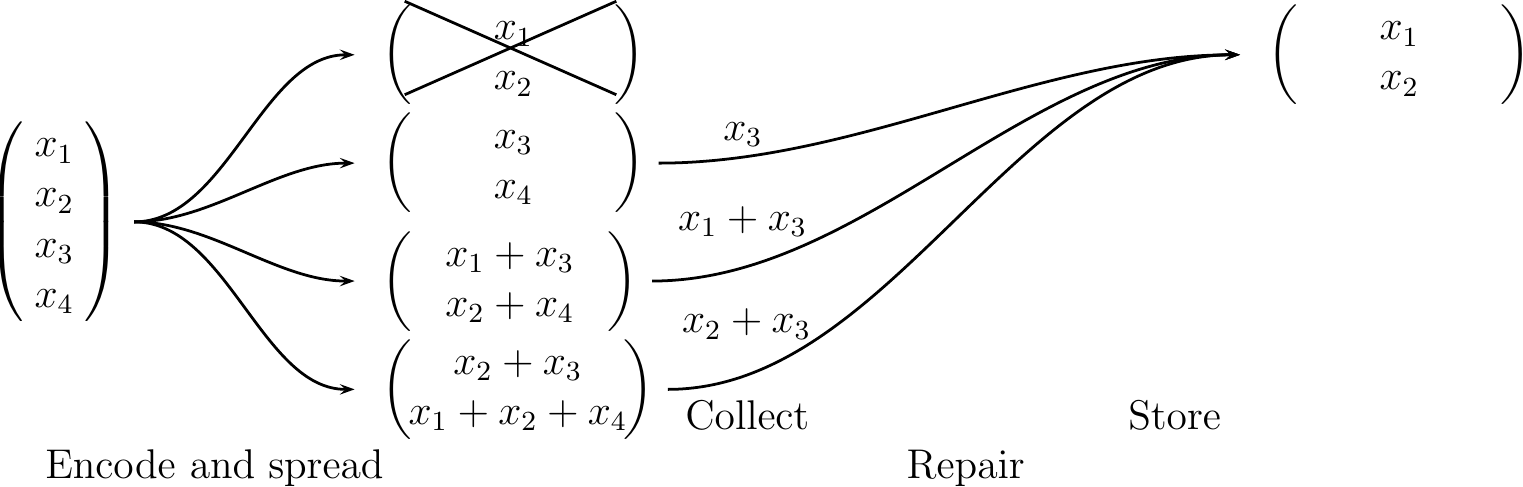}
        \label{fig:exact}
}
\caption{Regenerating codes are optimal with respect to storage and repair cost thanks to efficient repair methods.}
\label{fig:soa}
\end{figure*}

\section{Introduction}
As cloud-based solutions for backup and sharing are being offered to users, the amount of storage needed for cloud services keeps increasing. In order to lower the costs (\emph{e.g.}, hardware, energy) for operating such systems, it is important to rely on efficient paradigms. Currently, many systems still rely on well-proven replication~\cite{Ghemawat2003} to provide high availability from non-reliable devices. While easy to understand and implement, replication is far from being optimal with respect to the trade-off between storage and availability~\cite{Weatherspoon2002,Lin2004,Rodrigues2005}. Arguably, erasure correcting codes can significantly lower the amount of storage needed in data-centers. 

However, with classical erasure correcting codes (\emph{e.g.}, Reed-Solomon), reading data stored on an unavailable device (degraded read), or repairing after a permanent device failure generates many I/O operations and requires transferring a large amount of information over the network. In large-scale multi-site data-centers, the background network traffic due to degraded reads and repairs can become prohibitive for large amounts of data stored. In this paper, we focus on an attractive alternative, namely regenerating codes~\cite{Dimakis2010}, to lower such network costs.

Regenerating codes offer the same properties as erasure correcting codes with respect to storage and availability. Yet, as opposed to erasure correcting codes, regenerating codes significantly lower the network traffic upon repairs or degraded read. The seminal paper~\cite{Dimakis2010} on regenerating codes applies network coding to storage systems and defines the optimal trade-off between the amount of data which are stored and transferred. Regenerating codes,  designed to be as generic as possible, rely on many parameters, which are difficult to grasp in practice where device availability vary from one system to another. Moreover, many variants of regenerating codes exist (\emph{e.g.},~\cite{Ho2006, Suh2011, Rashmi2011, Cadambe2011, ElRouayheb2010}).

In order to help choose the right parameters and coding scheme, we make the following contributions:
\begin{itemize}
\item We study the influence of the various parameters at the system level, depending on storage device availability. We show that the optimum at device level does not always apply at system level. (Section \ref{sec:sys})
\item We compare the computational costs of various coding schemes for regenerating codes (random codes~\cite{Ho2006}, product-matrix codes~\cite{Rashmi2011}, and exact linear codes~\cite{Suh2011}) to the costs of classical erasure correcting codes (Reed-Solomon codes). (Section~\ref{sec:perf})
\end{itemize}

Previous practical work on regenerating codes focused either only on random codes~\cite{Duminuco2009} while we consider several other codes; or on a specific system with a specific code~\cite{Hu2011, Hu2012} while we study several codes and give conclusions that can be applied broadly.

\section{Model and Background}
We consider a system of $n$ devices connected by a network. The system stores files of size $\mathcal{M}$ that are immutable (\emph{i.e.}, data is appended to the system and once written cannot be modified, as in \cite{Calder2011}). Devices are available with a probability $p$ because of temporary disconnections (\emph{e.g.,} reboot, software upgrade,  short-term network disruption) or simply because they are overloaded.  

 When using erasure correcting codes, each file is divided into $k$ blocks and $n > k$ encoded blocks are produced so that any $k$ encoded blocks allow recovering the file. The file remains available as long as at least $k$ devices are available. Hence, the resulting system availability is $A=\sum_{i=k}^{n}{\binom{n}{i}p^i(1-p)^{n-i}}$. Whenever a block is permanently lost (\emph{e.g.}, disk crash, device replacement, long-term network disruption), a repair mechanism is used to regenerate it. This same repair mechanism is also used for degraded read (\emph{i.e.}, whenever a block is read but the device storing it is overloaded or temporarily unavailable). We assume that repairs (after a failure, or for a degraded read) are performed  according to a Poisson process with a rate of $\lambda$ repairs per day per device. 
 
The repair procedure in erasure correcting codes consists in contacting $k$ live devices, recovering the file and encoding it again to produce a new block. Since the entire file is read from disks and transferred over the network, this procedure has both high I/O costs, which can be reduced using specific codes~\cite{Huang2007,Papailiopoulos2012a,Prakash2012,Khan2012,Huang2012}, and high network cost, which can be reduced using regenerating codes~\cite{Dimakis2010}. In this paper, we focus on the latter codes reducing network costs.

\begin{figure*}[t]
\centering
\subfloat[MSR]{\centering%
	\includegraphics[width=0.48\linewidth]{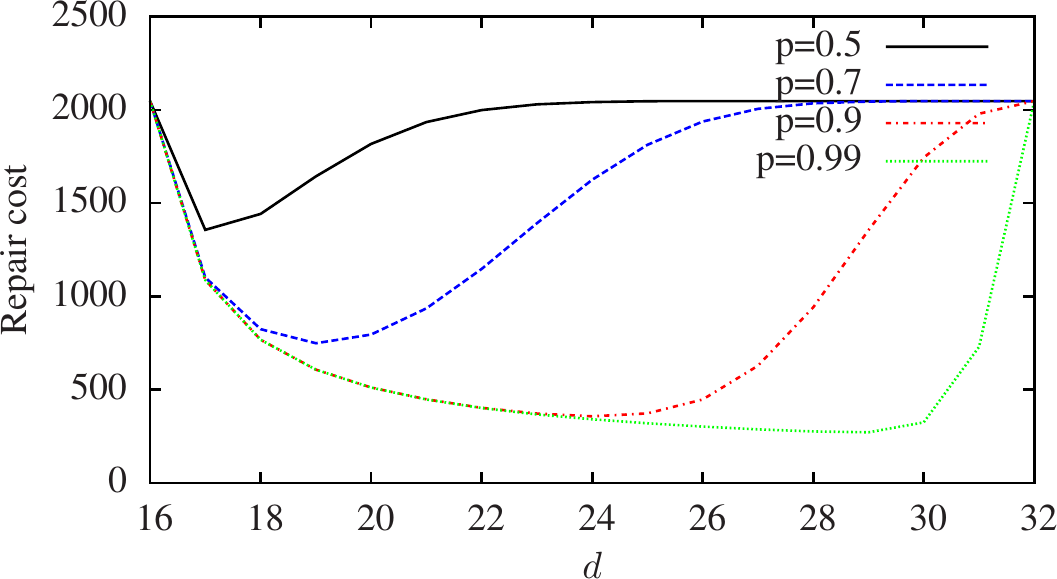}
	\label{sf:c_d_k_msr}
	}\hfill%
\subfloat[MBR]{\centering%
	\includegraphics[width=0.48\linewidth]{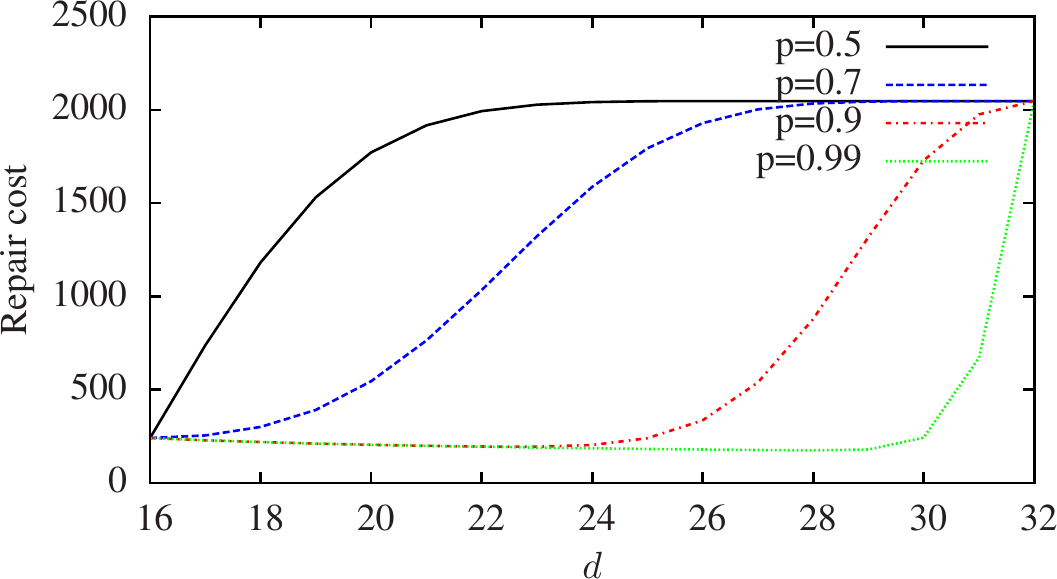}
	\label{sf:c_d_k_mbr}
	}
	\caption{System level repair cost as a function of $d$ for $k=16$ and $n=32$. The repair cost admits a minimum that is $\dopt=29 < n-1$ for MSR codes and $\dopt=28 < n-1$ for MBR codes.}
\label{fig:firstbis}
\end{figure*}

Regenerating codes apply network coding to storage systems to offer the best trade-off between network bandwidth repair cost $\gamma$ and storage cost $\alpha$. The file is divided into $k\Delta$ sub-blocks. These $k\Delta$ original sub-blocks are encoded into $n\alpha$ encoded sub-blocks which are then spread out across the $n$ devices (\emph{i.e.}, one group of $\alpha$ sub-blocks is stored on each device) so that contacting any $k$ devices allows recovering the file. Regenerating codes rely on an additional parameter $d$, which is the number of devices involved in a repair. Whenever a failure occurs, if the number of available devices is at least $d$, the following optimal repair method, shown in Figure~\ref{fig:exact}, can be used: \emph{(i)} the device being repaired fetches $\beta$ sub-blocks\footnote{In the rest of the paper, we will focus on scalar codes ($\beta=1$) for the sake of clarity and for they have lower computational costs. Vector codes ($\beta > 1$) have higher computational costs but are useful for reducing I/Os.} from each of $d$ available devices (thus $\gamma=d\beta$), \emph{(ii)} the device stores $\alpha$ sub-blocks computed from the $d\beta$ sub-blocks it received. 

Regenerating codes, as explained in~\cite{Dimakis2010}, can be parameterized by the values $\Delta$ and $\alpha$ to minimize either the storage (MSR, Minimum Storage Regenerating) or the bandwidth (\emph{i.e.}, network repair cost) (MBR, Minimum Bandwidth Regenerating). This trade-off is illustrated on Figure~\ref{fig:cr}. 

The seminal paper about regenerating codes relies on randomized code constructions, namely random linear network codes~\cite{Ho2006}. With such codes, the repaired data is not strictly equal to the lost data. As a first consequence, such codes cannot be maintained in a systematic form. Systematic codes  suppresses decoding costs and allows direct access to small parts of the file since the file can be read directly from the $k$ first blocks (or $k\alpha$ first sub-blocks) without decoding. As a second consequence, checking the integrity of such randomized codes requires to use complex techniques. In order to solve these two issues, it has been proposed to rely on exact regenerating codes~\cite{Dimakis2010b}. Various code constructions have been proposed; two of the most advanced ones were proposed by Suh~\emph{et al.}~\cite{Suh2011} and by Rashmi~\emph{et al.}~\cite{Rashmi2011}. In Section~\ref{sec:perf}, we will compare the computational costs of these schemes with the randomized code constructions~\cite{Ho2006} and regular Reed-Solomon erasure correcting codes.

Finally, in regenerating codes~\cite{Dimakis2010}, the number $d$ of devices to contact during repairs is chosen once for all and cannot be changed. Adaptive regenerating codes~\cite{NetCod2011} relax this constraint and allow $d$ to adapt to each repair. However, adaptive regenerating codes currently can only be implemented with random linear network codes, which have a high complexity and require costly schemes for integrity checking. In the following section, adaptive regenerating codes are included for they show the best achievable theoretical bound, yet further research is needed before they can be used in practice. To repair "static" regenerating codes when less than $d$ devices are available, $k$ available devices must be chosen and the repair must be carried by decoding the file before encoding it again leading to a cost $\gamma'=k\Delta$ as with regular erasure correcting codes. Hence, finding the right value for $d$ is important to avoid using this expensive repair by decoding method.



 In the rest of the paper, we study regenerating codes and put some emphasis on MSR codes, which offer the same trade-off between storage and availability as Reed-Solomon codes.

\section{System level analysis}
\label{sec:sys}
In this section, we consider several levels of device availability and study regenerating codes by performing an analysis at the system level since it matches the real costs observed. At the device level, the storage cost is $\alpha$ and the repair cost (\emph{i.e.}, network bandwidth) is $\gamma$. Hence, at the system level, the storage cost is $n\alpha$ and the repair cost is $\Gamma=n\lambda\gamma$.  The cost at the device level is known to decrease as $n$ increases. However, this conclusion does not apply at the system level. This section shows some interesting interactions between parameters. In the settings shown on the plots in the rest of the paper, we consider one file of size $\mathcal{M}=64 \mathrm{MB}$ and a repair rate of $\lambda=1$ per day per device. The repair cost is given in MB per day per file stored. This cost scales linearly with the number of files, the file size $\MM$, and the repair rate $\lambda$ (\emph{i.e.}, if 10 files are stored or if $\lambda=10$ repairs per day, the cost is 10 times higher) thus allowing extrapolating results.

\begin{figure*}[t]
\subfloat[MSR]{\centering%
                \includegraphics[width=0.48\linewidth]{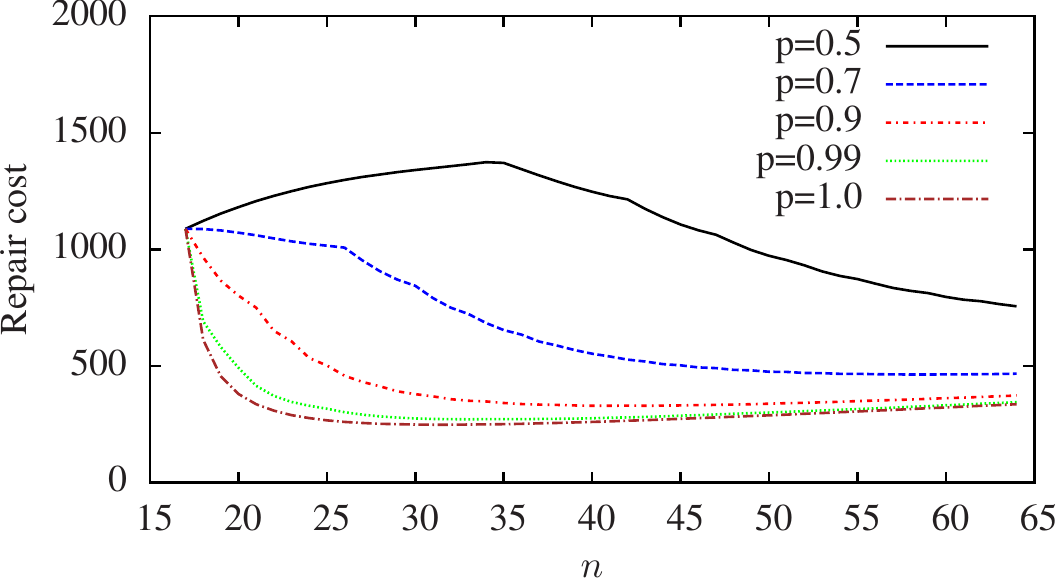}%
                \label{sf:bestn_msr}%
                }\hfill%
\subfloat[MBR]{\centering%
                \includegraphics[width=0.48\linewidth]{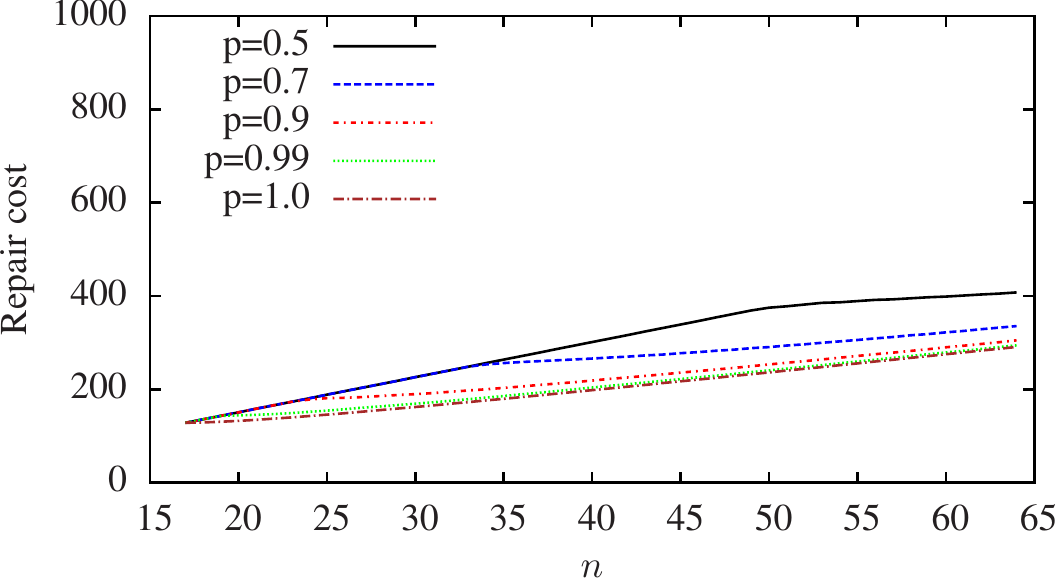}%
                \label{sf:bestn_mbr}%
                }%
 \caption{System level repair cost as a function of $n$ for $k=16$ and $d=\dopt$ where $\dopt$ is the best possible $d$ for a particular configuration $(n,k)$. For MSR codes, the optimal value depends on the device availability $p$. For MBR codes, the optimal value is $n=k+1$ with $d=k$.}
\label{fig:first}
\end{figure*}
\subsection{How many devices to repair from?}
\label{sec:31}
Let us consider that $k$ and $n$ are chosen to reach a given system availability~\cite{Weatherspoon2002,Lin2004,Rodrigues2005}. Regenerating codes require choosing an additional parameter $d$, which is the number of live devices contacted during a repair. Theoretical papers suggest that the best value for $d$ is $\dopt=n-1$. However, it turns out that this choice is not the best as soon as the device availability is $p<1$ as we explain in this section.

Suppose $X$ is a random variable describing the number of devices that are available. It can take values according to the following probability law
$$P(X=i)=\binom{n}{i}p^i(1-p)^{n-i}$$
Let $g$ be a function such that $g(i)$ is the cost of repair when $i$ devices are available. If $i \ge d$ devices are available, we repair using the optimal method
$$g(i)=\MB\frac{d}{\Delta}\;\; \textrm{ when } i \ge d$$  
Otherwise, we repair by decoding the whole file
$$g(i)=\MM\;\; \textrm{ when } i < d$$  
Hence, the expected repair cost is 
$$E[g(X)] = \sum_{i=k}^{n-1}P(X=i)g(i)$$
These repairs are performed at a rate $\lambda$ for each device. Hence, at the system level for the $n$ devices, the cost is 
$$n\lambda{}E[g(X)]$$
When less than $k$ devices are available\footnote{This case remains rare as the system availability $P(X \ge k)$ is high.}, the repair or the degraded read is simply delayed. To this end, we plot the system repair cost as 
$$\frac{n\lambda{}E[g(X)]}{P(k \le X \le n-1)}$$
which expand to
\begin{equation}
\frac{n\lambda{}\left(\sum_{i=k}^{d-1}{P(X=i)\MM}\, +\, \sum_{i=d}^{n-1}{P(X=i)\MB\frac{d}{\Delta}}\right)}{P(k \le X \le n-1)}
\label{eq:cost}
\end{equation}

\begin{figure*}[t]
\centering
\subfloat[optimal $n$ (MSR)]{\centering%
	      \includegraphics[width=0.24\linewidth]{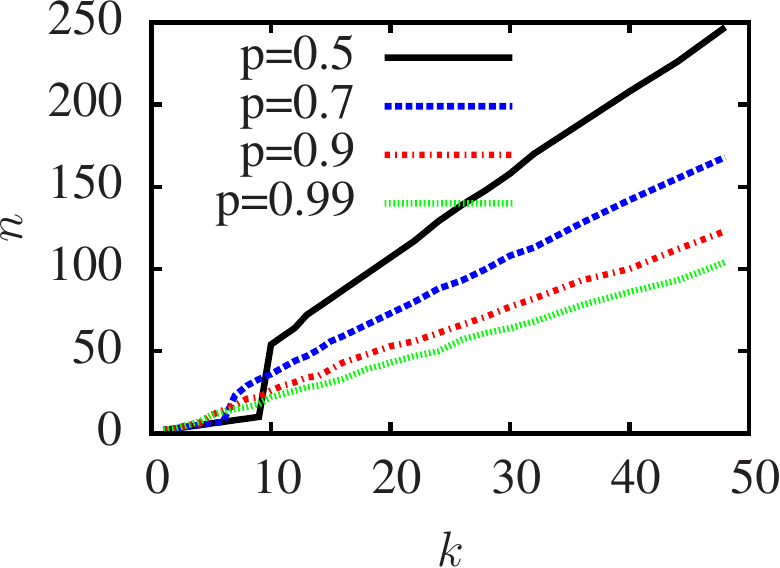}%
}\hfil
\subfloat[optimal $d$ (MSR)]{\centering%
	     \includegraphics[width=0.244\linewidth]{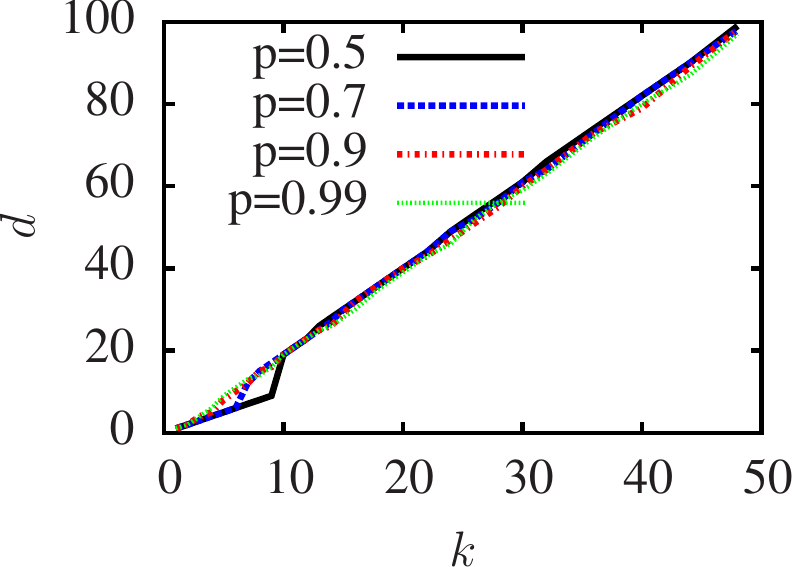}%
}\hfill
\subfloat[optimal $n$ (MBR)]{\centering%
	      \includegraphics[width=0.24\linewidth]{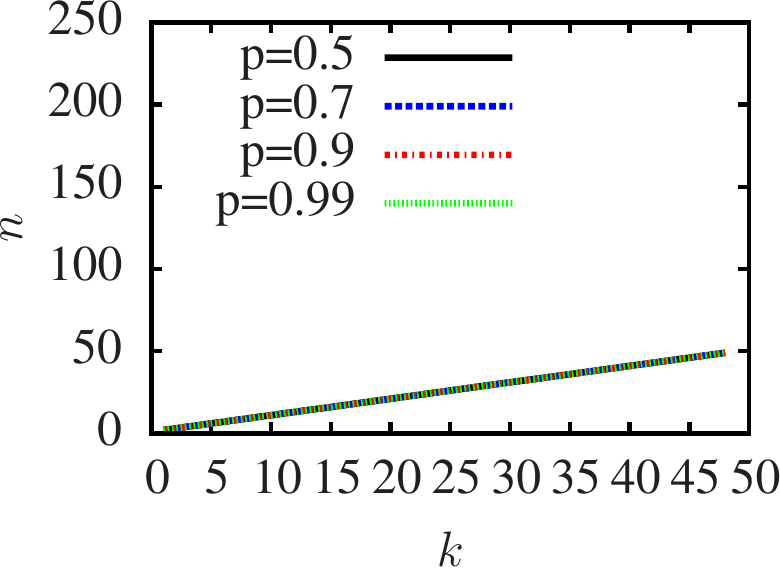}%
}\hfil
\subfloat[optimal $d$ (MBR)]{\centering%
	     \includegraphics[width=0.244\linewidth]{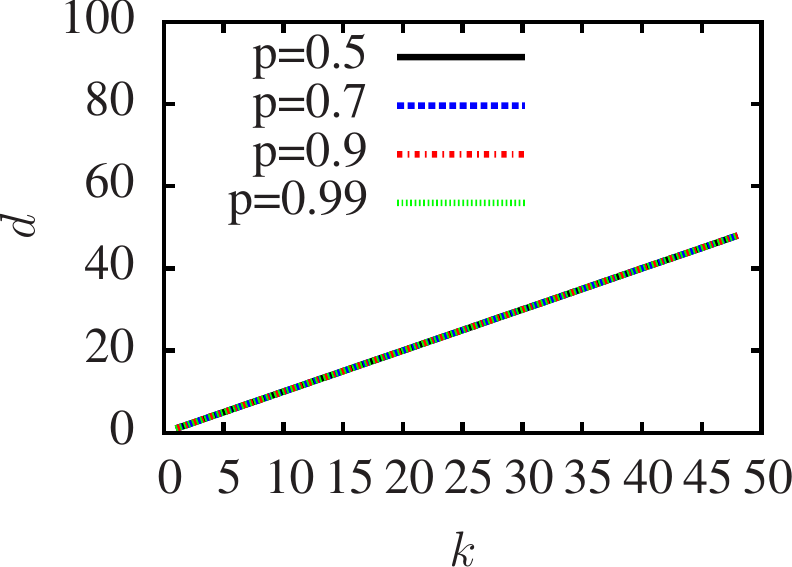}%
}
\caption{For MSR codes, the lowest repair cost is for $\nopt \approx k\times{}c(p)$ and $\dopt \approx 2k$. For MBR codes, the lowest repair cost is for $\nopt=k+1$ and $\dopt=k$.}
\label{fig:nkd}
\end{figure*}

According to theoretical studies~\cite{Dimakis2010},  a high value for $d$ helps reducing the cost of repairs at the device level. Yet, increasing $d$ also increases the probability that less than $d$ devices are available thus leading to more frequent repairs by decoding. Indeed, 2 devices being unavailable is much more frequent than 10 devices being unavailable at the same time. Hence, there should be an optimal value for $d$. For always available devices ($p=1$), it appears that $\dopt=n-1$, as stated in~\cite{Dimakis2010}.

In Figures~\ref{sf:c_d_k_msr} and~\ref{sf:c_d_k_mbr}, we consider a system relying on regenerating codes with $n=32$ and $k=16$. We consider various device availabilities $p$ and plot the system level cost as a function of $d$ as described in~\eqref{eq:cost}. We observe that the cost function admits an optimal value for $d$. For low to medium availabilities ($p=0.5$ to $p=0.9$), the optimal value for $d$ is rather low (much lower than $n-1$ value suggested by the literature~\cite{Dimakis2010}). For high availability $p=0.99$, the optimal value is close to $n-1$ but is still $\dopt=29$ ($\dopt=n-3$) for MSR codes, and $\dopt=28$ for MBR codes. Moreover, choosing $d=31$ instead of $d=\dopt$ when $p=0.99$ more than doubles the repair cost. Hence, as soon as devices are not perfectly available ($p \neq 1$), the designer must choose $d$ according to the device availability observed in the system to best leverage regenerating codes. 

When using MSR codes (Figure~\ref{sf:c_d_k_msr}), choosing a low value of $d$ is penalizing. Indeed, for $d=k$ or $d=k+1$, the cost is much higher than the optimal cost obtained when $d=\dopt$. For MBR codes (Figure~\ref{sf:c_d_k_mbr}), the behavior is completely different, and choosing a low value of $d$ (\emph{e.g.}, $d=k$) leads to a repair cost only slightly sub-optimal.

The repair cost for erasure correcting codes is the same as the cost for MSR codes with $d=k=16$. For high device availability ($p=0.99$), MSR codes with $d=\dopt$ offer an improvement by a factor of 10 over erasure correcting codes, and MBR codes offer an improvement by a factor of 12. For medium availability ($p=0.7$), with relatively low $\dopt=19$, MSR codes still offer an improvement by a factor of 3 over erasure correcting codes; and MBR codes offer an improvement by a factor of 10 with $\dopt=16$.

As explained, the system is rather sensitive to the choice of value $d$. This calls for codes where $d$ can be changed on the fly, namely adaptive regenerating codes~\cite{NetCod2011} (ARC), which are similar to MSR codes. These codes may seem more practical since they can self-adapt to the system, yet they currently lack practical code designs. Indeed, current code designs rely on randomized constructions rather than exact constructions, which have higher computational costs, and which require complex techniques to ensure integrity.  We discuss the potential of adaptive regenerating codes in more details in Section~\ref{sec:arc}

\subsection{How to choose the redundancy level?}
As with regular erasure correcting codes, the amount of redundancy must be chosen so that the  resulting availability $P(X \ge k) = \sum_{i=k}^{n}\binom{n}{i}p^i(1-p)^{n-i}$ is at least the desired availability $A$. This guides the values of the parameters $k$ and $n$ as studied in~\cite{Weatherspoon2002,Lin2004,Rodrigues2005}. Yet, with regenerating codes, it might be interesting to sacrifice storage efficiency in favor of repair cost by increasing $n$. Indeed, increasing $n$ allows to increase $d$ and hence to reduce repair cost. However, spreading the data across more devices increases the rate $n\lambda$ at which one of the $n$ devices may fail, thus increasing the number of repairs. 

\begin{figure*}
\subfloat[$k=16, n=32$  (ARC in bold, MSR in thin)]{\centering%
	     \includegraphics[width=0.48\linewidth]{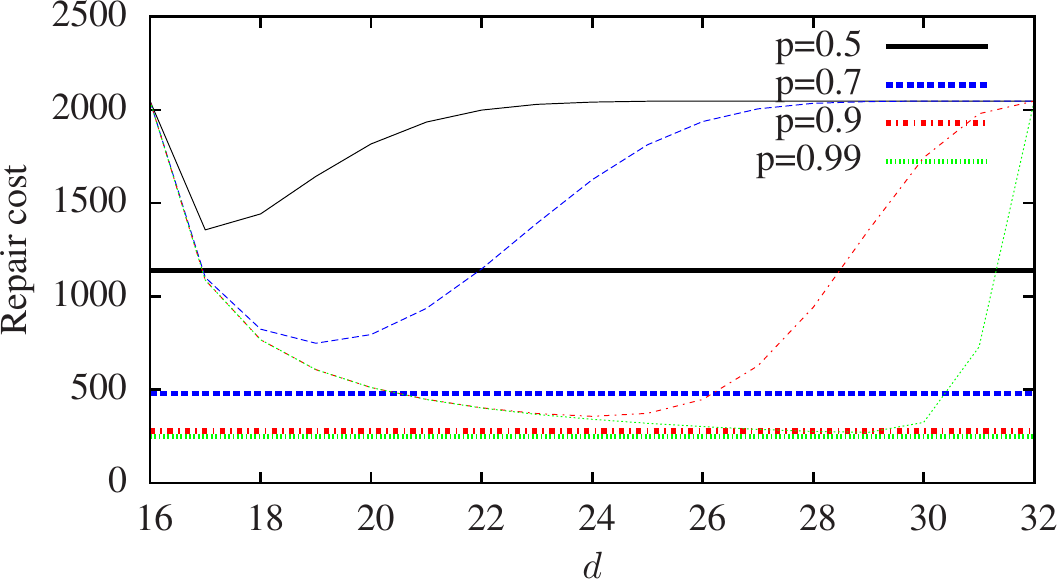}%
	     \label{sf:adapt}%
	     }\hfill%
\subfloat[Improvement obtained when using ARC in place of MSR codes]{\centering%
	     \includegraphics[width=0.48\linewidth]{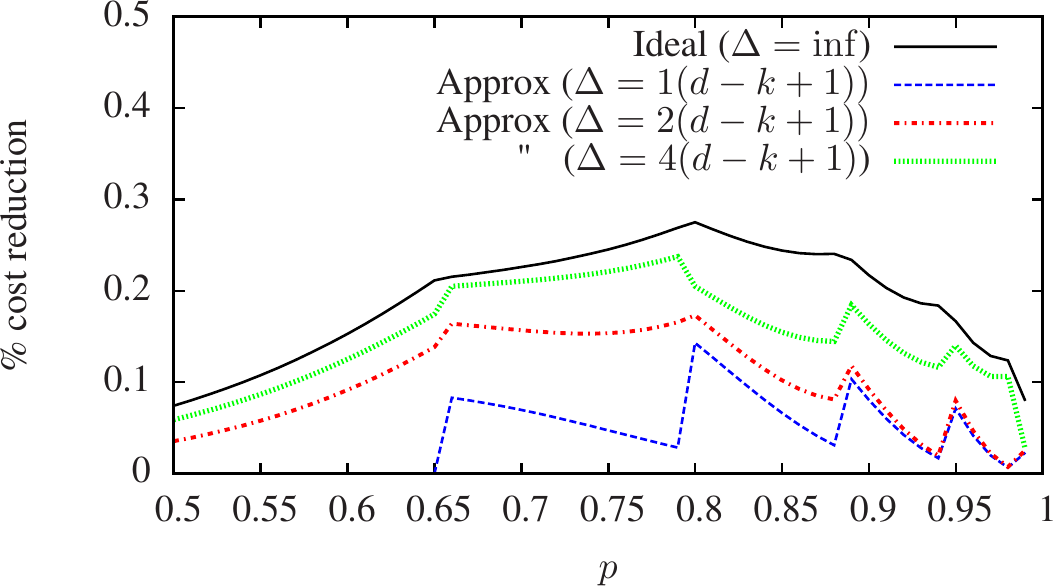}%
	     \label{sf:adapt2}%
	     }%
	     \caption{Adaptive regenerating codes lower the cost even when MSR are configured with the best value for $d$. The reduction in cost can be as high as 30\%. }
\end{figure*}

Let us consider that  $p=1$ and hence $d=n-1$. As initially observed by Dimakis \emph{et al.}~\cite{Dimakis2010}, selecting very large $n$ (and consequently allowing very large $d=n-1$) do not necessarily help in reducing costs. Indeed, for MSR codes, the system level repair cost $n\lambda\gamma$ admits a minimum at $\nopt={k+\sqrt{k^2-k}}$ (see~\cite{Dimakis2010}). When $k$ is large enough, $\nopt \approx 2k$. For MBR codes, the system level repair cost $n\lambda\gamma$ increases with $n$ ($n > k$) and thus admits a minimum at $\nopt=k+1$. 

We perform a similar study for $p < 1$ in order to minimize the repair cost, without taking into consideration availability requirements, which have been widely studied (\emph{e.g.}, \cite{Weatherspoon2002,Lin2004,Rodrigues2005}). Figures~\ref{sf:bestn_msr} and~\ref{sf:bestn_mbr} plot the value of $n$ which gives the lowest repair cost. For MSR codes, the optimal $\nopt$ depends on $p$ and is high. For MBR codes, the optimal value is $\nopt=k+1$. Indeed, as discussed previously in Section~\ref{sec:31}, the repair cost is only slightly reduced between $d=k+1$ and $d=\dopt$, so that increasing $n$ has a negative impact.

Figure~\ref{fig:nkd} shows the optimal value $\nopt$ and the corresponding value $\dopt$ for various values of $k$ and $p$. For MSR codes, we observe that $\dopt \approx 2k$ and $\nopt \approx k\times{}c(p)$ where $c(p)$ is a constant that depends only on $p$. The resulting $\nopt$ ensures that on average, the number of devices available for repair is approximately $\dopt$. For low values $(p,k)$, the best reported setting is $d=k, n=k+1$ which correspond to erasure correcting codes. This means that regenerating codes cannot operate efficiently for any value $(n,d)$ and erasure-correcting codes offer the lowest repair cost (but also a low reliability) for such $(p,k)$. However, slightly increasing $k$ is sufficient to leverage regenerating codes in spite of low device availabilities $p$. For MBR codes, the conclusion for $p<1$ is identical to the one given by theoretical analysis for $p=1$: the optimal value is $\dopt=k$ and $\nopt=k+1$ for all values $p$.

In summary, $k$ should be chosen to be sufficiently large for codes to be efficient, and $n$ should be chosen to guarantee the required availability, with criterion identical to the one applied when using regular erasure correcting codes (\emph{i.e.}, $n$ must be chosen so that the resulting availability is greater than or equal to the required availability $A$). 
$$
A \le \sum_{i=k}^{n}{\binom{n}{i}p^i(1-p)^{n-i}}
$$

Then, when using MSR codes, increasing $n$ should be considered as a way to reduce the repair cost at the price of some storage overhead. Once $n$ and $k$ are fixed, $d$ should be chosen carefully, depending on the device availability $p$ because an inappropriate $d$ negatively impacts the repair cost. However, predicting in advance the device availability $p$ is difficult, hence it would be interesting to have codes where the number of devices used for repair ($d$) can be changed at anytime. To this end, in the next section, we study adaptive regenerating codes, which do not require a fixed value $d$.


%
%
%
%
%


\subsection{How adaptive codes would perform ?}
\label{sec:arc}
\begin{figure}[!h]
	\centering
	\includegraphics[width=0.95\linewidth]{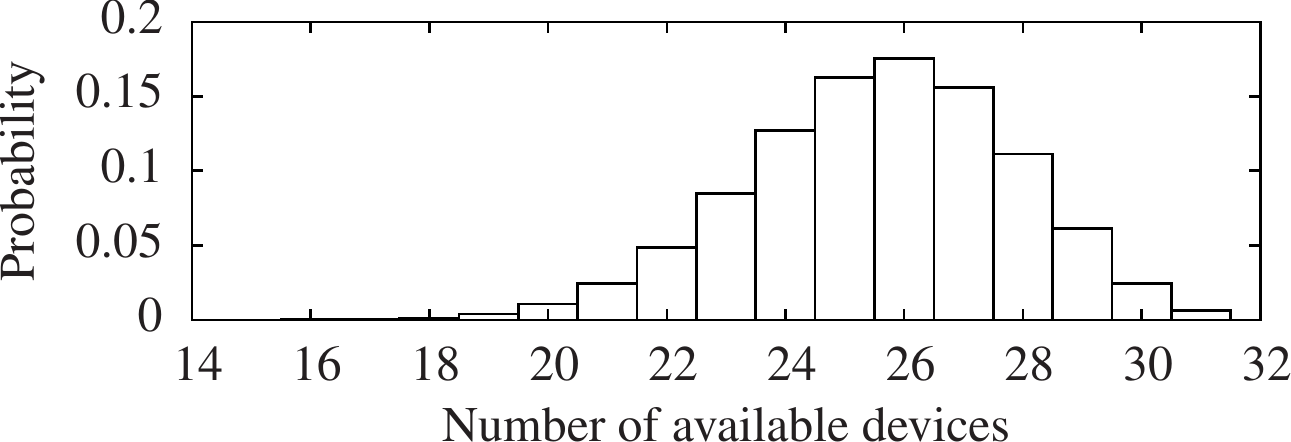}
	\caption{Probability that $x$ devices are available.}
	\label{fig:number}
\end{figure}

Regenerating codes~\cite{Dimakis2010} rely on the assumption that the parameter $d$ is fixed once for all. However, in practice the number of devices that are available in a system is spread around a mean value. If we consider that one device has failed in a system consisting of $n$ independent devices, the number of remaining available devices follows the probability law shown in Figure~\ref{fig:number}. Hence, it is of interest to use codes which can accommodate any value of $d$ rather than codes which can only accommodate one specific value of $d$. To this end, we study adaptive regenerating codes that have been defined in~\cite{NetCod2011}. Adaptive regenerating codes have the same storage and bandwidth requirements as MSR codes\footnote{Adaptive regenerating codes relies on the fact that, at the minimum storage point, different repairs do not depend on each other. However, at the minimum bandwidth point, as a repair depends heavily on previous repairs, it is not possible to adapt the value $d$. Hence, there exist no MBR-like adaptive regenerating codes.}. In this study, to allow a fair comparison, we do not rely on the ability to repair multiple failures in a coordinated way as it would artificially lower the cost for adaptive regenerating codes in some cases.

\begin{figure*}[t]
\centering
\subfloat{\centering%
                \includegraphics[width=0.45\linewidth]{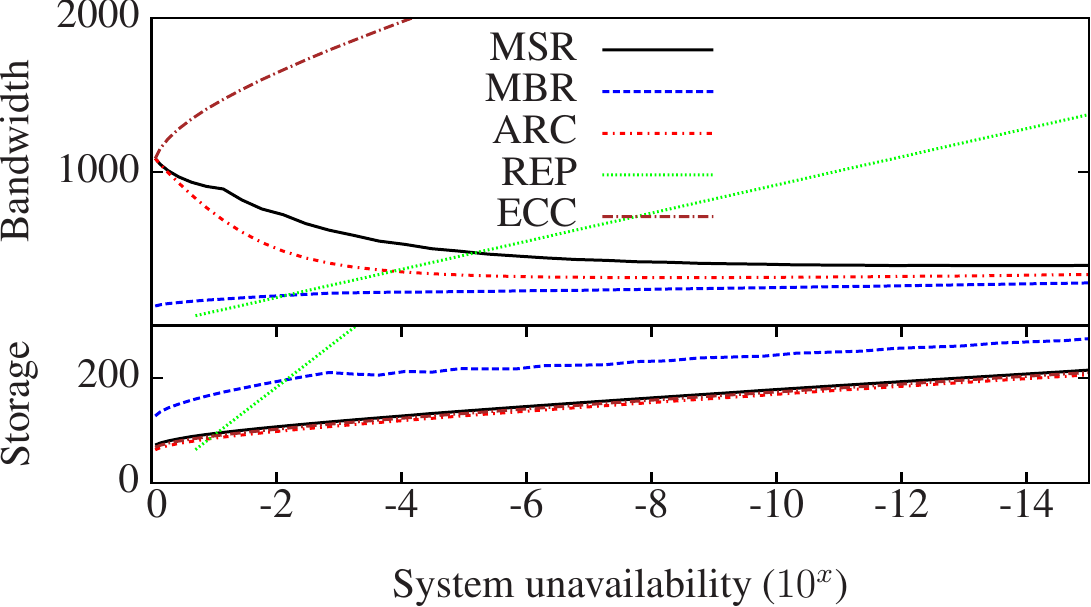}%
                \label{sf:multi}%
}
\hfill
\subfloat{\centering%
                \includegraphics[width=0.45\linewidth]{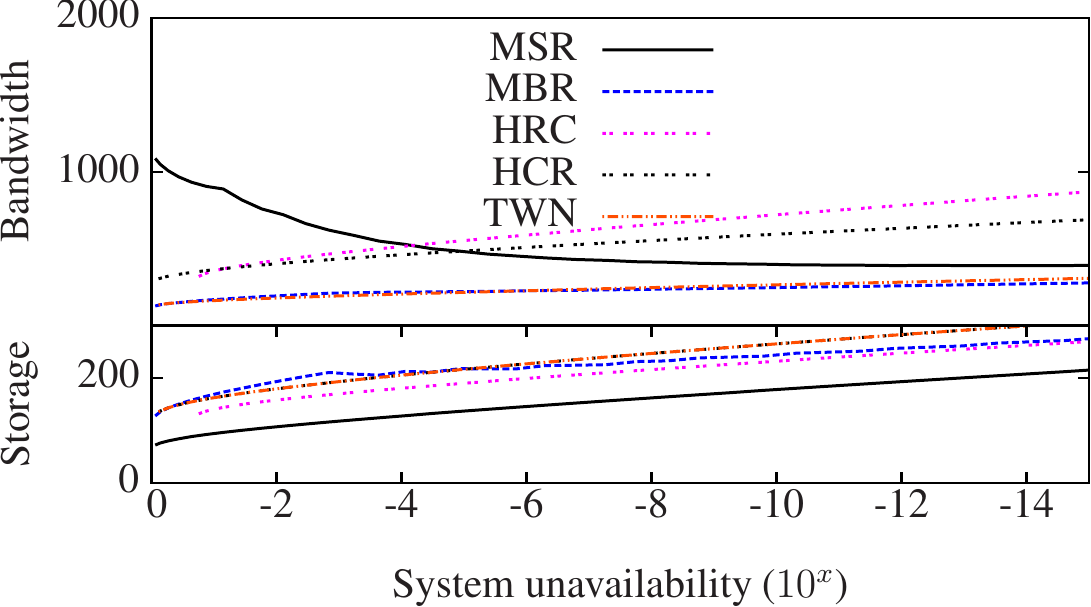}%
                \label{sf:multi2}%
}
\caption{Storage and bandwidth tradeoff for various method, for various system availability given a device availability of $0.8$.}
\label{fig:multi_main}
\end{figure*}

Figure~\ref{sf:adapt} plots repair costs for optimal adaptive regenerating codes (ARC) in bold lines alongside regular codes in thin lines\footnote{When using adaptive regenerating codes~\cite{NetCod2011}, repairs are performed independently to compare to MSR and MBR on a fair basis (\emph{i.e.}, no coordinated multiple repairs).}.  First, adaptive regenerating codes are simpler to deploy since they require not setting a value $d$ that depends on $p$. An interesting observation is that adaptive regenerating codes, have a lower repair cost than MSR codes with $d$ set to the optimal value. 

Figure~\ref{sf:adapt2} plots the improvement obtained when using adaptive regenerating codes instead of MSR codes depending on the device availability $p$. Adaptive regenerating codes ($\Delta=\infty$) achieve up to 25\% repair cost reduction for systems with a device availability of around $80\%$ (\emph{i.e.} devices are overloaded or unavailable 20\% of the time). The periodicity that can be observed in the plot comes from the fact that for a given value $k$, the values $\beta$ that occur frequently are not the same for all values $p$, and that for some value $p$, these values $\beta$ are closer to an entire multiple of $\frac{1}{\Delta}$ than for other values of $p$.

When using adaptive regenerating codes, various amounts of data must be sent. For example, consider a code where $k=2$ and using $\beta$ as defined in~\cite{NetCod2011}. If 3 devices are available, each available device sends $\beta=\frac{\MM}{2}\frac{1}{d-k+1}=\frac{\MM}{2}\frac{1}{2}$ bytes of data to the device being repaired. If 4 devices are available, each available device sends  $\beta=\frac{\MM}{2}\frac{1}{3}$. This is achieved by splitting the file into $k\Delta=2\times{}(2\times{}3)$ sub-blocks, and sending 3 sub-blocks if 3 devices are available (\emph{i.e.}, $\beta=\frac{\MM}{2}\frac{3}{6}$) and 2 sub-blocks if 4 devices are available (\emph{i.e.}, $\beta=\frac{\MM}{2}\frac{2}{6}$). Hence, to be able to repair optimally using a wide range of values $d \in \{k, n-1\}$, we must have a large $\Delta=\textrm{lcm}({n-k,1})$ (see~\cite{Kermarrec2010INRIA} for more details). This tends to increase the computational cost when compared to MSR codes where $\Delta=d-k+1$. 

To address the issue of having too many sub-blocks, an alternative is to take a smaller $\Delta$. In this case, live devices send just enough sub-blocks so that the amount of data sent is greater than or equal to the amount needed (\emph{e.g.}, sending $\beta=\frac{\MM}{2}\frac{2}{5}$ when $\beta=\frac{\MM}{2}\frac{1}{3}$ is needed). This gives an approximate version of adaptive regenerating codes. In Figure~\ref{sf:adapt2}, we explore such approximate adaptive regenerating codes. We observe that taking $\Delta=d-k+1$ performs poorly. However, a limited $\Delta=4(d-k+1)$ is sufficient to approach the savings obtained with ideal adaptive regenerating codes.

To summarize, adaptive regenerating codes can lower costs over regular regenerating codes. The savings can reach 25\% in the case we studied, and they do not require to set a precise value for $d$ thus simplifying system configuration. However, currently, the only known coding schemes are \emph{(i)} randomized schemes~\cite{Kermarrec2010INRIA}, or \emph{(ii)} deterministic schemes limited to $k=2$~\cite{LeScouarnec2012a}. Hence, further research is needed to design exact adaptive regenerating codes for use in practical systems.

\subsection{Which codes to choose ?}
Replications policies differ in their specification in a way that inhibits comparison. Indeed, they often rely on $(n,k)$ notation, which indicates that the $k$ original blocks are encoded and stored as $n$ blocks. However, the property of the schemes in term of reliability differ and such notation does not give any indication on the availability obtained. For example three codes with same $(n,k)$ such as MSR codes~\cite{Dimakis2010}, MBR codes~\cite{Dimakis2010} and Twin codes~\cite{Rashmi2011a} have different reliability guarantees, and different storage overhead. In order to compare codes on a fair basis, we propose to compare them according to the desired system unavailability, their storage overhead, and their repair cost.

In Figure~\ref{sf:multi}, we compare MSR codes~\cite{Dimakis2010}, MBR codes~\cite{Dimakis2010}, adaptive regenerating codes~\cite{NetCod2011} (ARC), erasure correcting codes (ECC) (\emph{e.g.}, Reed-Solomon codes) and replication (REP). Overall, regenerating codes outperform erasure correcting codes and replication, except when the system is unavailable more than 1\% of the time, which is an unrealistic setting. Hence, regenerating codes are appealing for replacing both regular erasure correcting codes and replication. 

Figure~\ref{sf:multi} also allows us to compare MBR and MSR codes. Indeed, MBR codes rely on increasing the storage overhead (by storing more data on each device) to reduce bandwidth consumption for repairs. By increasing $n$ beyond the minimum value needed to ensure the desired availability, MSR also allows us to increase the storage overhead to reduce the bandwidth consumption. If we set a storage budget of 200, we can obtain a repair bandwidth of 214 for MBR codes (and a system unavailability of $10^{-2}$). For the same budget, we can obtain a repair bandwidth 400 of for MSR codes (and a system unavailability of $10^{-13}$). For a fixed system availability, MBR consumes twice the storage space needed for MSR; but MBR significantly reduce the bandwidth consumption, especially for system with a low availability. The system designer should choose MBR if saving network bandwidth in a system offering a low availability is the priority, otherwise MSR codes are a better option.

Finally, ARC\footnote{Again, for the comparison to be fair, repairs are performed independently without relying on coordinated multiple repairs capability of ARC, which would reduce bandwidth consumption even more.} approaches MBR with respect to bandwidth without sacrificing the storage efficiency (as efficient as MSR) thus significantly improving the performance for a wide range of desired system availabilities. 

\begin{figure*}[t]
\centering
\subfloat[Time for encoding a file in seconds]{\centering
                \includegraphics[width=0.30\linewidth]{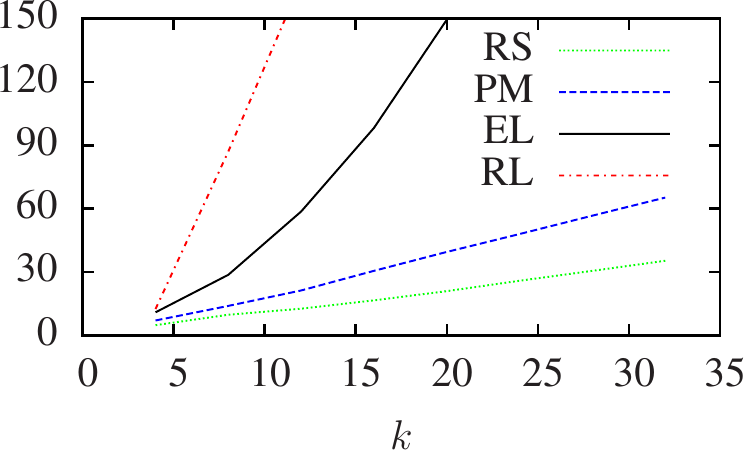}%
                \label{fig:perf1a}
                }\hfil
\subfloat[Time for repairing a lost device in seconds]{\centering
	     \includegraphics[width=0.30\linewidth]{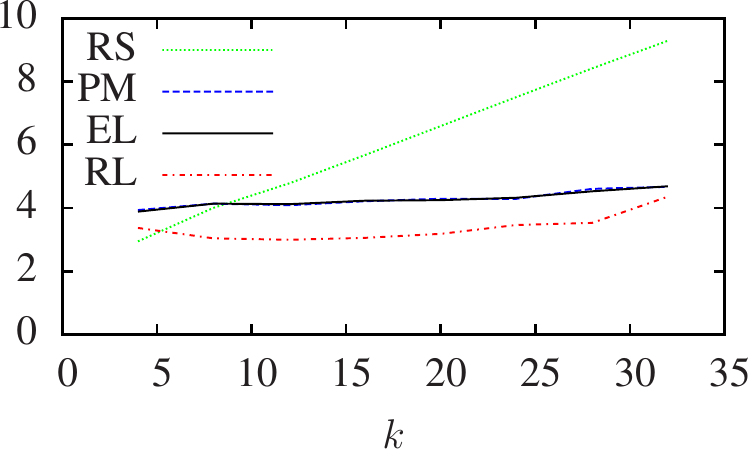}%
	                     \label{fig:perf1b}
	     }\hfil
\subfloat[Time for decoding a file in seconds]{\centering
                \includegraphics[width=0.30\linewidth]{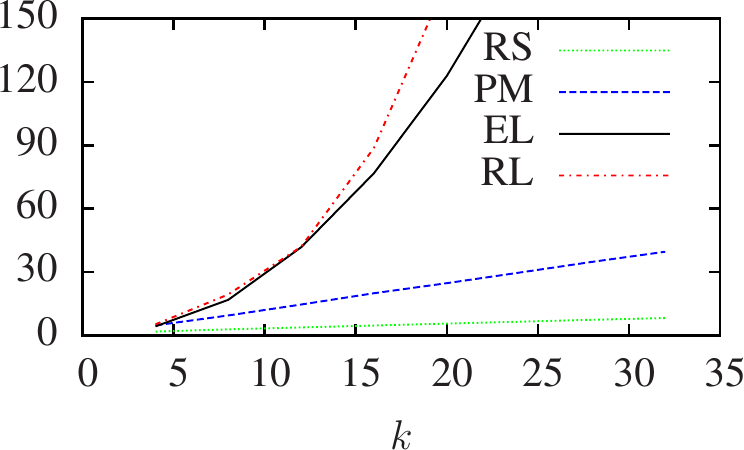}%
                                \label{fig:perf1c}
}
\caption{Performance for $\mathcal{M}=16 \mathrm{MB}$.  Reed-Solomon decode the file and encode the lost block at each repair.} 
\label{fig:perf1}
\end{figure*}

In Figure~\ref{sf:multi2}, we leverage our comparison framework to explore some other codes such as Twin Codes~\cite{Rashmi2011a} (TWN), as well as hybrid coding schemes applying erasure correcting codes on top of replication (HRC), or replication on top of erasure correcting codes (HCR). Hybrid schemes are outperformed by MBR codes for both storage and bandwidth efficiency. They are also outperformed by MSR codes when the system availability is high. Twin Codes~\cite{Rashmi2011a} are codes that have a low storage overhead ($\MB$ per device), and optimal repair cost ($\MB$ per failed device) but that jeopardize the reliability to achieve this (\emph{i.e.}, the file cannot be recovered from any $k$ devices out of $n$ but only from some subsets of $k$ devices). The comparison method we proposed, based on the availability rather than simply on the $(n,k)$ parameters, allows us to compare them to regenerating codes on a fair basis. Interestingly, as shown on Figure~\ref{sf:multi2}, Twin Codes have costs very similar to MBR codes and are much more efficient than the hybrid schemes.

\begin{figure*}[t]
\centering
\subfloat[Time for encoding a file in seconds]{\centering
                \includegraphics[width=0.30\linewidth]{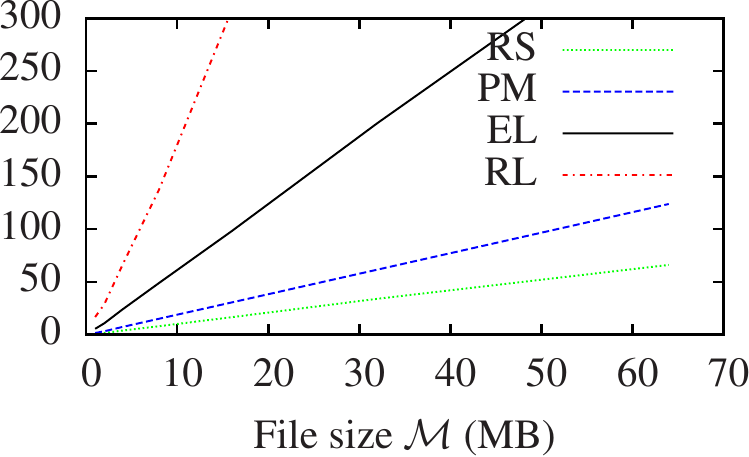}%
                }\hfil
\subfloat[Time for repairing a lost device in seconds]{\centering
\label{fig:perf3b}
	     \includegraphics[width=0.30\linewidth]{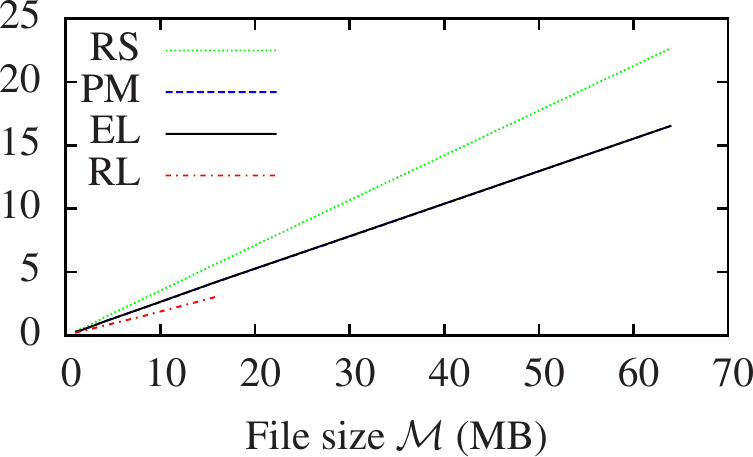}%
	     }\hfil
\subfloat[Time for decoding a file in seconds]{\centering
                \includegraphics[width=0.30\linewidth]{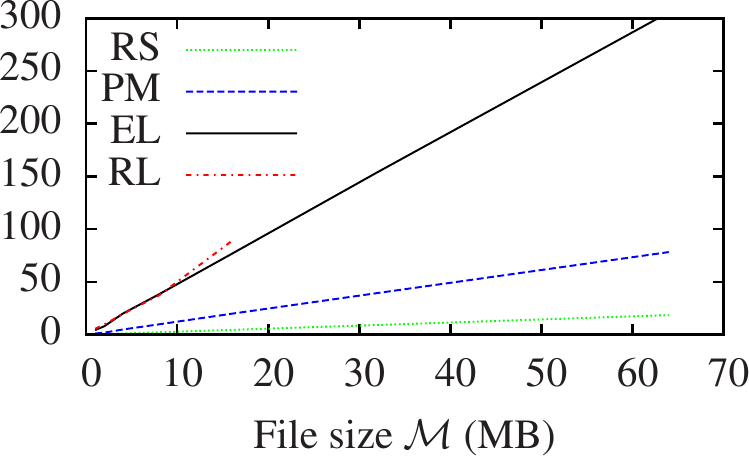}%
}
\caption{Performance for $k=16$.  Reed-Solomon decode the file and encode the lost block at each repair.} 
\label{fig:perf2}
\end{figure*}

\section{Computational Performance}
\label{sec:perf}
Various code designs exist to implement regenerating codes. These codes rely on various algorithms and data structures, leading to different  costs when implemented. In this section, we consider both CPU processing and memory costs. We focus on codes that offer the same properties as regular erasure correcting codes (\emph{e.g.,} Reed-Solomon codes) in term of storage cost and reliability, because they offer a drop-in replacement for erasure correcting codes in existing systems. Hence, we compare the following most significant such codes, which are Reed-Solomon codes for erasure correcting codes, and random linear network codes~\cite{Ho2006},  exact linear codes~\cite{Suh2011} and product-matrix codes~\cite{Rashmi2011} for MSR regenerating codes. Our observations also give some hints at the performance of some other codes (\emph{e.g.}, product-matrix MBR codes rely on the same algorithms as product-matrix MSR codes, and hybrid codes rely on the same algorithms as Reed-Solomon codes for the encoding and decoding processes).

\subsection{Memory costs}
We briefly study the costs associated with the data structures needed to be loaded in memory for encoding and decoding. 

Reed-Solomon codes are linear codes which rely on an $n\times{}k$ matrix containing elements of $q$ bits. For $k=16, n=32, q=16$, the resulting encoding matrice is of size 1 Kbytes. 

Linear network codes (random linear network codes~\cite{Ho2006} or exact linear codes~\cite{Suh2011}) rely on an  $k\alpha{}\times{}n\Delta$ (approximately $k^2\times{}nk$ for MSR codes). For $k=16, n=32, q=16$, the resulting encoding matrice is of size 256 Kbytes. 

The product-matrix codes~\cite{Rashmi2011} rely on a more compact scheme and the encoding matrix is of size $n\times{}2\alpha$ (approximately $n\times{}2k$ for MSR codes). For $k=16, n=32, q=16$, the resulting encoding matrice is only 2 Kbytes. 

Overall, when considering 64 MB data blocks (typical in cloud storage systems), the memory requirements for these matrices is negligible. Moreover, apart for random linear network codes, which are non-deterministic codes, the encoding matrices, which are the same for all files, are created using a deterministic process and need not be stored since they can be re-created on the fly when needed with a low computational cost. As a consequence, memory costs, even if higher for regenerating codes than for erasure correcting codes, are only a minor issue and are not relevant for choosing one particular code design over another. However, as we will explain hereafter, the processing costs are a true limitation and vary greatly from one code design to another.

\begin{figure*}[t]
\centering
\subfloat[Time for encoding a file in seconds ($\MM=16 \mathrm{MB}$) (all systematic codes on the left, comparison of the best codes with their non systematic version on the right)]{\centering
                \includegraphics[width=0.30\linewidth]{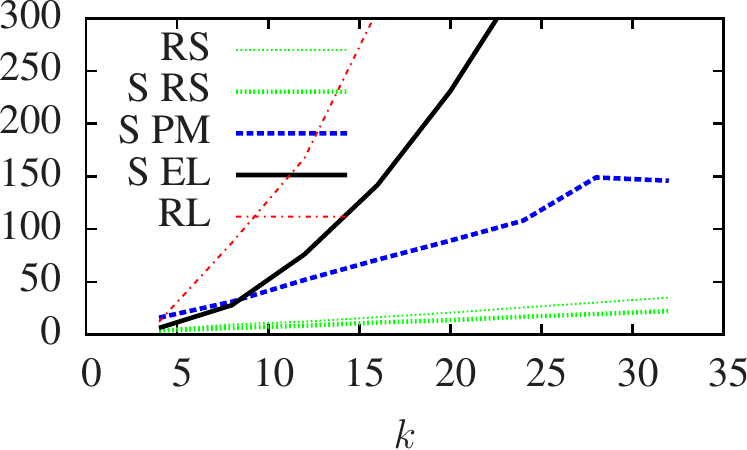}%
                \hfil%
                \includegraphics[width=0.30\linewidth]{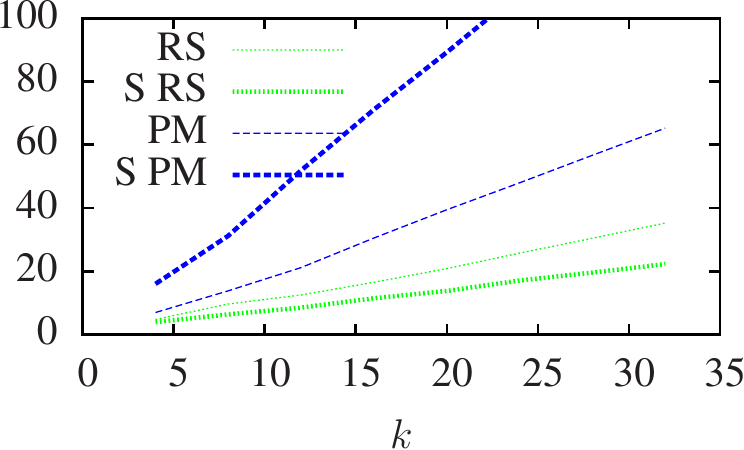}%
}\hfil\hfil%
\subfloat[Time for encoding a file in seconds ($k=16$)]{\centering
	     \includegraphics[width=0.30\linewidth]{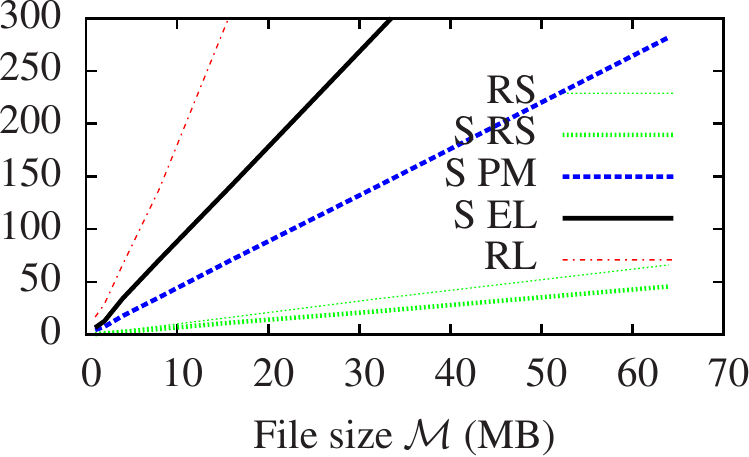}%
	     }%
\caption{Performance for systematic codes. The repair procedure remains unchanged (see Figure~\ref{fig:perf1}). No decoding is needed when accessing the file.} 
\label{fig:perf3}
\end{figure*}
\subsection{CPU costs}
To compare the processing costs, we implemented in Java several MSR codes. All codes implementations have similar levels of optimization and rely on a log-table based finite field implementation\footnote{Computational costs using multiplication-table based finite field implementations are 30\% lower but with such finite field implementation $k$ is limited because of constraints on field size.}.  We ran these mono-threaded implementations on a Pentium E2200.

We implemented \emph{(i)} random linear network codes (RL)~\cite{Ho2006}, \emph{(ii)} exact linear codes (EL)~\cite{Suh2011}\footnote{We use a linear implementation of product-matrix codes~\cite{Rashmi2011}, leading to constructions similar to~\cite{Suh2011}.}, \emph{(iii)} product-matrix codes that use a compact representation of codes with efficient encoding and decoding algorithms (PM)~\cite{Rashmi2011}, and \emph{(iv)} Reed-Solomon erasure correcting codes (RS)~\cite{Reed1960}.

Regular erasure correcting codes (\emph{e.g.}, Reed-Solomon) involve linear operations on matrices of size $k\times{}k$. Such operations have a reasonable complexity $\Omega{(k^2)}$. However, regenerating codes involve the same linear operations but on matrices of size $k\alpha{}\times{}n\Delta$ (approximately $2k^2\times{}k^2$ when $n=2k$ and $d=n-1$). Consequently, naive linear implementations (EL, RL)~\cite{Ho2006,Duminuco2009,Suh2011} suffer from a complexity of $\Omega{(k^4)}$ and have high computational costs even for low values of $k$. It can be observed that Product-Matrix codes~\cite{Rashmi2011} (PM) rely on efficient algorithms departing from classical linear approaches thus lowering costs as shown here. 

Figure~\ref{fig:perf1} shows the time needed to process a file of size $\MM=16 \mathrm{MB}$  depending on the parameter $k$. When considering the encoding (Figure~\ref{fig:perf1a}), all regenerating codes (PM, EL, RL) perform worse than regular erasure correcting codes (RS). However, it is interesting to notice that product-matrix codes (PM) clearly outperform regular linear regenerating codes codes (EL, RL). The two linear regenerating codes rely on the same encoding and decoding algorithm, yet, exact linear codes (EL) clearly outperform random linear network codes (RL). Indeed, the encoding matrix of exact linear codes (EL) is much sparser than the one of random linear network codes (RL). Overall, these results are consistent with the asymptotic complexities discussed in the previous paragraph. 

When considering the repair time  (Figure~\ref{fig:perf1b}), all regenerating codes (PM, EL and RL) exhibit similar costs, with a slight advantage for random linear network codes (RL) because their randomized repair procedure is simpler. Reed-Solomon codes, whose repair procedure rely on a costly decoding proccess followed by an encoding proccess suffer a cost which increases with $k$ and which is higher than the cost of all regenerating codes schemes. As a consequence, when repair operations are frequent, it can be more interesting from a computational point of view to use regenerating codes even if encoding and decoding are more costly.

When considering the decoding time, Reed-Solomon (RS) codes have a low cost thanks to a specific decoding algorithm and small encoding matrices. Product-matrix codes (PM) have a reasonable cost that is much lower than the other regenerating codes (EL, RL). Indeed, product-matrix codes use a specific decoding algorithm much more efficient than the algorithm used for other regenerating codes (EL, RL).

Figure~\ref{fig:perf2} shows that the encoding time, the repair time and the decoding time scale linearly with the file size $\MM$, and confirms the relatively good performance of product matrix codes (PM), and the poor performance of linear codes (EL, RL) when compared to classical erasure correcting codes (RS).

Figure~\ref{fig:perf3} studies the additional cost of relying on systematic codes. Systematic codes are interesting because the $k$ first devices store non-encoded data. As a result, when performing a read on such codes, no decoding is needed (\emph{i.e.}, the decoding cost is null).  Reed-Solomon codes use a specific encoding matrix that encodes directly to a systematic form. As this matrix is sparser than the matrix for the non-systematic version, the systematic Reed-Solomon (S RS)~\cite{Roth1985} is faster than the non-systematic version (RS). On the contrary exact linear regenerating codes (S EL)~\cite{Suh2011} use a matrix that is costly to create and that is denser than the original one leading to higher costs. Finally, product matrix codes rely on a pre-coding step that is performed before the encoding step. This precoding step uses an algorithm similar to the decoding algorithm and as such increases the costs: the systematic product matrix codes (S PM)~\cite{Rashmi2011} are more costly than the non-systematic ones (PM).

Hence, product-matrix codes are good candidates for replacing Reed-Solomon codes in practical systems. Their impact on memory and CPU remains limited. As shown in Figure~\ref{fig:perf3b}, their non-systematic form only doubles the encoding costs and quadruples the decoding costs when compared to non-systematic Reed-Solomon codes. Their systematic form increases the encoding costs by a factor of 7 when compared to systematic Reed-Solomon codes. Systematic product-matrix codes should be preferred when data is read more frequently than it is written, otherwise  non-systematic codes are more efficient.




 
For MBR codes, which are not the focus of this paper, Fractional Repetition Codes~\cite{ElRouayheb2010}, not implemented in this benchmark, perform very well since they rely on an efficient systematic pre-code (\emph{e.g.} Reed-Solomon) to produce encoded sub-blocks that are then replicated on several devices. Repairs and reads are performed using simple transfers without any computation. The two other implementations of MBR codes are random linear network codes based~\cite{Ho2006}  or product-matrix codes based~\cite{Rashmi2011}, and since they use the same algorithms at the MBR and the MSR point, they will behave similarly to their MSR implementation (PM and RL).

 \section{Conclusion and Discussion} 
We study the impact of various parameters of regenerating codes since they can have a significant impact at the system scale.
First, despite common belief, $\dopt$ is not necessarily $n-1$ and instead should be carefully tuned due to its high impact on the repair method efficiency at the system level. Even for high device availability $p=0.99$ where $d=n-1$ seems reasonable, yielding improvement by a factor of 2 when choosing $d=n-3$. Second, the lowest repair cost for MSR codes is obtained with $d=2k$ and $n=k\times{}c(p)$ where $c(p)$ is a constant depending on the device availability; the lowest repair cost for MBR codes is obtained with $d=k$ and $n=k+1$. Third, if sufficient system availability is achieved with a rather low $n$, designers who need to further lower repair cost should favor MBR over using MSR with artificially increased $n$ because MBR achieves lower repair costs at the system scale. Finally, since adaptive regenerating codes theoretically outperform MSR at the system level when looking at the amount of data transferred; providing exact code designs for ARC can be a promising theoretical research area that would allow ARC to be implemented in practical systems. 

We also study the computational costs associated with the various coding schemes available. We show that the additional cost of using regenerating codes (product-matrix codes~\cite{Rashmi2011}) is reasonable. We have shown that non-systematic product-matrix codes outperform other linear code designs and only double (resp. quadruple) the cost of encoding (resp. decoding) when compared to  Reed-Solomon codes. Systematic product-matrix codes increase by a factor of 7 the cost of encoding when compared to systematic Reed-Solomon codes. Hence product-matrix codes keep computational costs within reasonable values given the savings in term of the amounts of network bandwidth they allow to achieve. As a perspective, to lower systematic codes computational costs, it would be interesting to design product-matrix like codes that encode directly to a systematic form without requiring a pre-processing step.

Interesting perspectives and ongoing theoretical work for applying regenerating codes to practical system concern the minimization of I/O operations jointly with network repair cost and storage cost. Indeed, we focused on regular regenerating codes minimizing the network cost. In some systems, minimizing the I/O operations (disk reads) is equally important. Regenerating codes are not incompatible with this consideration and this research subject is active. Recent work has optimized I/O cost by relying either on specific coding schemes~\cite{Cadambe2011,ElRouayheb2010} or considering variations of regenerating codes~\cite{Kiani2011,Papailiopoulos2012}. Also, reducing I/O is an interesting application for coordinated regenerating codes~\cite{NetCod2011} that support repairing multiple devices at once. Indeed, if instead of performing $t$ successive repairs, $t$ repairs are slightly delayed and performed at once in a coordinated way, the I/O operations for repairs are factored out thus reducing the overall I/O operations by a factor $t$ without jeopardizing the optimality with respect to network.%

\section*{Acknowledgment}
This study was partially funded by the ODISEA (Open Distributed Networked Storage Architecture) collaborative project from the competitiveness clusters System@tic and Images \& Réseaux. 
 
\bibliographystyle{abbrv}
\bibliography{hs}%
 \end{document}